\documentclass[12pt]{article}
\usepackage[top=1in, bottom=1in, left=1in, right=1in]{geometry}
\usepackage[english]{babel}
\usepackage{atbegshi,cite}
\usepackage{amsmath,amssymb,amsbsy,amstext, amsthm, simplewick}
\usepackage{hyperref}
\usepackage{graphicx}
\usepackage{amsfonts}
\usepackage[small]{caption}
\usepackage{upgreek}
\usepackage[titletoc]{appendix}
\usepackage{setspace}
\usepackage{color}

\newcommand{\newc}{\newcommand}
\newc{\fpi}{f_{\pi}}
\newc{\etap}{\eta^{\prime}}
\newc{\llll}{\langle\lambda\lambda\rangle}
\newc{\FFd}{F^a\tilde F^a}
\newc{\qbar}{{\overline q}}
\newc{\TR}{{\rm Tr}}
\newc{\Kahler}{K\"ahler }
\newc{\Zbb}{{\mathbb Z}}
\newc{\Rt}{{\mathbb R}^3}
\newc{\Rf}{{\mathbb R}^4}
\newc{\So}{{\mathbb S}^1}
\newc{\zt}{{\mathbb Z}_2}
\newc{\RtSo}{{\mathbb R}^3\times{\mathbb S}^1}
\newc{\scriminus}{{\cal I}^-}
\newc{\scriplus}{{\cal I}^+}
\newc{\mpl}{M_p}
\newc{\Ricci}{\mathcal{R}}
\newc{\bv}{\phi}
\newc{\calU}{{\cal U}}
\newc{\calK}{K}
\newc{\calUi}{{\cal U}^{-1}}
\newc{\calG}{{\cal G}}
\newc{\calO}{{\cal O}}
\newc{\calOb}{{\cal O}^\dagger}
\newc{\hphi}{{\hat\phi}}

\newcommand{\stat}[1]{\hat{#1}}

\theoremstyle{plain}
\theoremstyle{plain} 
\theoremstyle{plain} 
\theoremstyle{plain}
\theoremstyle{plain}
\theoremstyle{plain}

\renewcommand{\title}[1]{{\Large\bf\flushleft{#1}}\vspace*{3ex}\\}
\renewcommand{\author}[2]{{\noindent\hspace*{2.5em}\large#1}
                     \footnote{Electronic mail: $\mathtt{#2}$}\\}
\renewcommand{\d}{\mathrm{d}}

\begin{document}
\begin{titlepage}
\begin{flushright}
{\large 
~\\
}
\end{flushright}

\vskip 2.2cm

\begin{center}

{\large \bf Gravitational Instabilities and Censorship \\
of Large Scalar Field Excursions}

\vskip 1.4cm

{ Patrick Draper$^{(a)}$ and Szilard Farkas}
\\
\vskip 1cm
{$^{(a)}$ Department of Physics, University of Illinois, Urbana, IL 61801}\\
\vspace{0.3cm}
\vskip 4pt

\vskip 1.5cm

\begin{abstract}

Large, localized variations of light scalar fields tend to collapse into black holes,  dynamically ``censoring" distant points in field space. We show that in some cases, large scalar excursions in asymptotically flat spacetimes can be UV-completed by smooth Kaluza-Klein bubble geometries, appearing to circumvent 4d censorship arguments. However, these spacetimes also exhibit classical instabilities related to the collapse or expansion of a bubble of nothing, providing a different censorship mechanism. We show that the Kerr family of static KK bubbles, which gives rise to an infinite scalar excursion upon dimensional reduction, is classically unstable. We construct a family of initial data in which the static bubbles sit at a local maximum of the energy, and we give a general argument that such a property indeed indicates mechanical instability in gravity. We also discuss the behavior of wound strings near a bubble, a local probe of the large traversal through moduli space.

\end{abstract}

\end{center}

\vskip 1.0 cm

\end{titlepage}
\setcounter{footnote}{0} 
\setcounter{page}{1}
\setcounter{section}{0} \setcounter{subsection}{0}
\setcounter{subsubsection}{0}
\setcounter{figure}{0}



\section{Introduction}

Given a theory with a long flat direction in field space, can a local observer access distant points in that space? Simple arguments suggest that the answer is often no. Consider a free scalar field undergoing a large excursion $\Delta\varphi$ in a spatial region of size $R$, with typical field gradients $\sim \Delta\varphi/R$. Then the energy density can be made arbitrarily small by taking $R$ large. However, when coupled to gravity, $R$ is less than the Schwarzschild radius of the experiment when $\Delta\varphi\gtrsim M_p$, independent of $R$. In other words, if one tries to set up a localized transplanckian field excursion, the experiment may collapse into a black hole.\footnote{This observation has been made a number of times~\cite{tbanks}, and related observations appear in~\cite{Banks:2003vp,Banks:2004im}.}  

A more detailed analysis was performed in~\cite{nicolis}. It was argued in~\cite{nicolis} that the variation of a massless scalar outside static, spherically symmetric, asymptotically flat, nonsingular sources in 4d gravity is bounded by a number of order one in Planck units. It is natural to ask more precisely under what circumstances gravity censors large scalar excursions\footnote{A different variety of censorship was observed in~\cite{ArkaniHamed:2007js}, in which it was noted that in some string theory compactifications, Euclidean wormhole solutions can be found along which moduli undergo a transplanckian excursion. \cite{ArkaniHamed:2007js} showed that such wormholes do not contribute to the path integral via duality arguments. Transplanckian censorship associated with large-$f$ axion excursions around cosmic strings was discussed in~\cite{draperetal}. For infinite strings, large axion excursions are not strictly censored. However, the spacetime becomes highly dynamical, bearing little resemblance to a cosmic string, and the excursion takes an exponentially long time to probe. Around finite loops of string, other arguments suggest censorship occurs~\cite{draperetal}.}, and whether the mechanism is always collapse into a black hole.

We will show that some of the 4d configurations used in the analysis of~\cite{nicolis} can be  UV-completed by static Kaluza-Klein (KK)  geometries. These solutions are part of a larger family of KK bubbles of nothing (BON) based on the Euclidean Kerr solution. All of the bubble geometries are singularity- and horizon-free in 5d, yet  give rise to an infinite  excursion of the KK scalar upon dimensional reduction. These examples thus appear to weaken the bound on $\Delta\varphi$. Is there still a notion of  censorship?

One observation is that curvature scales near the large scalar excursion are of order the KK radius, so that only the full higher-dimensional description is useful in this region. In other words, the excursion cannot be resolved without discovering that the theory is five dimensional. One might conclude that there is not really a transplanckian scalar to censor.

However, there is another property of the static, asymptotically flat KK bubbles  that suggests a stronger form of censorship: they are classically unstable. The instability is well-known for the Euclidean Schwarzschild bubble~\cite{grossperryyaffe} and we will show that the entire Kerr family is unstable. These instabilities may be anticipated by a mechanical analogy. Tunneling events in flat KK space studied in~\cite{BON,brillhorowitz} must occur under a potential barrier. The static, classical bubbles sit at a stationary point in this potential, which is not a minimum if the potential has only one extremum along a curve connecting flat space to a nucleating bubble. A static bubble at such a locus is analogous to a ball at the top of a hill, and the resulting classical instability effectively censors the large traversal in moduli space.

There is likely some relationship between transplanckian censorship and various conjectures about the swampland~\cite{swampland2,Obied:2018sgi,Agrawal:2018own} and the Weak Gravity Conjecture (WGC)~\cite{wgc}. The study of inflationary models with large scalar field excursions in time~\cite{extranatural} and the difficulties associated with finding long flat directions in string theory~\cite{banksdinefoxgorbatov} originally motivated the WGC, and the implications of the WGC for scalar field ranges / large moduli spaces is an active area of study~\cite{Brown:2015iha,Brown:2015lia,Heidenreich:2015nta,Heidenreich:2015wga,Klaewer:2016kiy,Palti:2017elp,Hebecker:2017uix,Heidenreich:2017sim,Lust:2017wrl,Hebecker:2017wsu,Heidenreich:2018kpg,Grimm:2018ohb,Cheung:2018cwt,Reece:2018zvv,Draper:2018lyw}. There is also evidence that the WGC is related to cosmic censorship~\cite{Crisford:2017gsb}. The precise connection with dynamical instances of transplanckian censorship like those we will discuss remains tantalizing but incompletely understood.

This work is organized as follows. In Sec.~\ref{sec:dimred} we  review the 4d bound on scalar traversals obtained in~\cite{nicolis}. We show that some of the 4d configurations considered in~\cite{nicolis} can be identified with smooth KK bubble solutions, and we note that a larger class of KK spacetimes produces large $\Delta\varphi$ upon dimensional reduction. In Sec.~\ref{sec:EKN} we show that all of these bubbles are classically unstable. We identify a 1-parameter set of 4d Euclidean Kerr-Newman metrics that provide a family of initial data connecting the static bubbles to the nucleation point of instantons described in~\cite{BON,dowkeretal}. Within this family, both flat space and the nucleating bubbles lie at zero energy, while the static bubbles appear at a local maximum of the energy, indicating mechanical instability. Sec.~\ref{sec:generalinstab} provides the technical argument that ``local maximum of the energy $\Rightarrow$ instability" for the static bubbles, and outlines a more general set of conditions under which this is true in gravity. In Sec.~\ref{sec:string}, we discuss how wound strings probe the local value of the KK scalar, treating the bubbles as quasistatic backgrounds. In Sec.~\ref{sec:disc} we summarize and conclude.

 Our discussion concerns solutions of classical KK theory. In supersymmetric KK theories, the radion may remain massless at the quantum level, but the solutions we will discuss no longer exist due to the fermion boundary conditions. In nonsupersymmetric theories, Casimir energies lift the radion, and fluxes and other objects are required to stabilize it. We will assume that the classical solutions with an exact moduli space are reasonable approximations to solutions in at least some more-complete theories with stabilized radion.\footnote{For example, it was observed in~\cite{Dine:2004uw} that BON instantons persist in the presence of simple stabilizing potentials.} (If this assumption is false, our discussion can be truncated at the introduction. But in this case it would appear that large scalar excursions are even more difficult to arrange in asymptotically flat KK spacetimes.)

\section{$\Delta\varphi$ and KK bubbles}
\label{sec:dimred}
\subsection{4d bounds on  $\Delta\varphi$}
Ref.~\cite{nicolis} obtained a bound on massless, spherically symmetric scalar field excursions in 4d general relativity. We briefly sketch the argument underlying the bound. 
A family of static, spherically symmetric solutions to 4d gravity+minimally coupled scalar was discovered by Buchdahl~\cite{buchdahl}:
\begin{align}
ds^2=-&f^{\beta}dt^2+f^{-\beta}dr^2+r^2 f^{1-\beta} d\theta^2+r^2 f^{1-\beta} \sin^2(\theta)d\phi^2\;, \nonumber\\
\varphi&=\sqrt{\frac{1}{16\pi}\left(1-\beta^2\right)}\log(f)\;,\;\;\;\;\;\;\;\;f\equiv 1-2  m/r\;.
\label{eq:buchdahl}
\end{align}
These solutions exhibit naked singularities at $r=2m$. If the solutions are cut off before reaching the singularity,~\cite{nicolis} argued that the cutoff should occur before the ADM mass is saturated by the exterior space,
\begin{align}
4\pi \int^\infty_{R_0}\rho(R) R^2 d R \leq M_{ADM}=\beta m\;,
\label{eq:adm}
\end{align}
working in coordinates $R(r)=r f^{\frac{1-\beta}{2}}$ where the angular part of the metric takes a canonical form. In these coordinates the singularity is at $R=0$ and we place a cutoff at $R_0$. The scalar energy density is
\begin{align}
\rho=\frac{1}{2}g^{RR}(\partial_R\phi)^2\;.
\end{align}
Eq.~(\ref{eq:adm}) yields a constraint on the scalar field excursion outside $R_0$, parametrically of order 
\begin{align}
\Delta\phi\lesssim1
\label{eq:phibound}
\end{align}
 in Planck units~\cite{nicolis}.

ADM masses of spherically symmetric spacetimes like~(\ref{eq:buchdahl}) are given by 
a positive volume integral over all of space and, in cases where $g_{RR}\rightarrow 0$ at $R=0$, a negative contribution at the singularity,
\begin{align}
M_{ADM}=\lim_{R_0\rightarrow 0}\left( 4\pi \int^\infty_{R_0}\rho(R) R^2 dR -\frac{R_0}{g_{RR}(R_0)}\right)\;.
\label{eq:admfull}
\end{align}
In the configurations~(\ref{eq:buchdahl}), the singularity in the integral at small radius is cancelled by the singularity in the surface term, yielding a finite ADM mass $m\beta$. Therefore, to establish a bound of the form~(\ref{eq:phibound}), it is important that the 4d curvature singularities imply a finite cutoff $R_0$, below which is it assumed that $g_{RR}(0)$ is nonzero in the UV completion, so that the surface term cannot make a (negative) contribution.

\subsection{Buchdahl spacetimes from KK bubbles}
In some cases, instead of cutting off the curvature singularity with a finite-radius source, the metric+scalar~(\ref{eq:buchdahl}) can be smoothly completed in Kaluza-Klein theory. In these cases, from the point of view of the 5d theory, singularities in the 4d curvatures are canceled by the KK scalar. Lacking singularities, the KK solutions do not require a cutoff, and so the bound of~\cite{nicolis} cannot be directly applied. 

The relevant KK solution was constructed in~\cite{GP}  by adding a Lorentzian time direction to 4d Euclidean Schwarzschild (ES) gravitational instanton~\cite{gibbonshawking}. We refer to this solution as the ES bubble. The line element is 
\begin{align}
ds^2=-dt^2+f^{-1} dr^2+&r^2 d\theta^2+r^2 \sin^2(\theta)d\phi^2 +f dx_5^2\;.\label{eq:KKbubble}
\end{align}
The radial coordinate runs from $r_H$ to $\infty$, where
\begin{align}
r_H=2m\;.
\end{align}
Absence of a conical singularity near $r_H$ requires  periodicity $x^5\in [0,8\pi m)$. The spacetime then ends in a smooth cap at $r_H$, marking the wall of a bubble of nothing. 

The metric can be parametrized as 
\begin{align}
g_{AB}=\left(\begin{matrix}
G_{\mu\nu}&0\\
0& V
\end{matrix}
\right)
=\sigma^{-1/3}\left(\begin{matrix}
g_{\mu\nu}&0\\
0& \sigma
\end{matrix}
\right)\;,
\label{eq:metparam}
\end{align}
suitable for dimensional reduction along $x_5$. The second form  yields the 4d Einstein frame,
\begin{align}
S=\frac{1}{16\pi}&\int d^4x \sqrt{-g} \left(-\Ricci_4-\frac{1}{6}g^{\mu\nu}\frac{\partial_\mu \sigma \partial_\nu\sigma}{\sigma^2}\right)\;\nonumber\\
=&\int d^4x \sqrt{-g} \left(-\frac{1}{16\pi}\Ricci_4-\frac{1}{2}\partial_\mu \varphi \partial^\mu\varphi\right)\;
\end{align}
where $\sigma=e^{\sqrt{48\pi}\varphi}$. 
The ES bubble has the 4d form
\begin{align}
g_{\mu\nu}dx^\mu dx^\nu=-f^{1/2}dt^2+f^{-1/2}dr^2&+r^2 f^{1/2} d\theta^2+r^2 f^{1/2} \sin^2(\theta)d\phi^2\;, \nonumber\\
\sigma&=f^{3/2}\;.
\label{eq:ESfourd}
\end{align}
We see that $\varphi\sim\log(1-r_H/r)$ is  a canonical, minimally coupled scalar diverging logarithmically near $r_H$, matching one of the 4d solutions of Eq.~(\ref{eq:buchdahl}),
\begin{align}
{\rm ES~bubble}\Longleftrightarrow{\rm ~Buchdahl~solution},~\beta=1/2.
\end{align}
 The ES bubble is thus a nonsingular 5d realization of a 4d scalar with a large, localized, static $\Delta\varphi$.

Similar relationships of this type have also been noted in the literature. Kastor and Traschen found that 4D cosmological spacetimes related to the Buchdahl spacetimes may  be obtained from dimensional reduction~\cite{Kastor:2016cqs}, while Garriga~\cite{Garriga:1998ri} and Brown and Dahlen~\cite{Brown:2011gt} have shown the singular Hawking-Turok instanton~\cite{Hawking:1998bn} is related by dimensional reduction to the nonsingular 5d Witten BON instanton.

\subsection{More general KK bubbles}
In bubble spacetimes, the KK scalar diverges where the circle radius goes to zero. Such points are fixed points of the Killing vector generating translations around the circle, and in general may either be isolated points (nuts) or form surfaces (bolts)~\cite{gibbonshawking}. One can thus obtain more general 4d spacetimes with large, localized excursions of a minimally coupled scalar modulus by dimensionally reducing KK solutions with nuts and bolts. Nut charge is associated with KK monopole number~\cite{GP,sorkin}. We will consider only asymptotically flat spacetimes with bolts, in which the large excursion occurs on an extended surface.

The simplest family of this type, which includes the ES bubble, is obtained by adding a Lorentzian time direction to the 4d Euclidean Kerr (EK) metric. The latter is obtained by analytic continuation of the Kerr metric ($t\rightarrow ix_5$, $a\rightarrow ia$) and a twisted periodic identification of coordinates. (These spacetimes were called ``KK dipoles" in~\cite{GP}; we will refer to them as EK bubbles.) The line element is
\begin{align}
ds^2=-dt^2+\Sigma^{-1}\big[\Delta(dx_5&+a\sin^2\theta d\phi)^2+\sin^2\theta((r^2-a^2)d\phi-adx_5)^2\big]\nonumber\\
&+\Sigma(dr^2/\Delta+d\theta^2)\;,\nonumber\\
\nonumber\\
\Delta= r^2-2m&r-a^2\;,\;\;\;\;\;\;\Sigma=r^2-a^2\cos^2\theta\;.
\label{eq:EKline}
\end{align}
There is an apparent singularity at 
\begin{align}
r_H=m+\sqrt{m^2+a^2}
\end{align}
which is also a fixed surface of the Killing vector~\cite{gibbonshawking}
\begin{align}
\ell=\partial_{x_5}+\Omega \partial_\phi\;,\;\;\;\;\Omega=\frac{a}{r_H^2-a^2}\;.
\end{align}
It is convenient to change coordinates to $\tilde\phi=\phi-\Omega x_5$, where the Killing vector is simply
\begin{align}
\ell=\partial_{x_5}\;.
\end{align}
Examining the metric near $r_H$, one finds that the conical singularity can be removed with the periodicity
\begin{align}
x_5\sim x_5+2\pi\gamma\;,\;\;\;\;\;\gamma=\frac{r_H^2-a^2}{\sqrt{m^2+a^2}}
\end{align}
at fixed $\tilde\phi$. In the original coordinates, we have
\begin{align}
(x_5,\,\phi)&\sim \left(x_5+2\pi\gamma,\,\phi+2\pi \Omega\gamma\right)
\label{eq:twisted}
\end{align}
and $r\geq r_H$. The spacetime is asymptotically flat with a nonstandard KK identification, and contains a bubble of nothing with radius $r_H$. In the limit $a\rightarrow 0$, $\Omega\rightarrow0$ and we recover the ES bubble of the previous section with the standard periodicities. 

Twisted periodicities are associated with magnetic flux in the reduced theory~\cite{mag1,mag2}. Reducing on $x_5$ at fixed $\tilde\phi$, the magnetic field in cylindrical coordinates $(\rho,\tilde\phi,z)$ behaves at large $z$ as
\begin{align}
&B(\rho=0,z\gg r_H) = 2\Omega\hat{\mathbf z}\;,\nonumber\\
&B(\rho\gg r_H,z\gg r_H) \rightarrow 0\;.
\label{eq:mag}
\end{align}

At large distances, the proper radius of the KK circle grows linearly as a result of the twisted identification. In cylindrical coordinates, 
\begin{align}
g_{55}\sim1+\Omega^2\rho^2\;,\;\;\;\;\;\; r\gg r_H\;.
\end{align}
For the purposes of KK reduction, spacetimes with twisted identifications should therefore be thought of as an approximation with a long-distance cutoff, $\rho\ll1/\Omega$. This point of view was emphasized in~\cite{dowkeretal}, with which we will make further contact below. 

Furthermore, since the surface $r=r_H$ is a spherical bolt for $\ell$, dimensional reduction gives rise to a divergence in the KK scalar on the bubble. Reducing on $x_5$ in the $\tilde\phi$ coordinates and expanding near $r_H$, we obtain
\begin{align}
\sigma\sim (r-r_H)^{3/2}\;,\;\;\;\;\;\; r-r_H\ll 1
\end{align}
leading again to a logarithmically divergence of the canonically normalized, minimally coupled scalar $\varphi$.

\section{Instabilities}
\label{sec:EKN}

In the dimensionally reduced bubble spacetimes, the KK scalar only becomes large in a region close to $r_H$. In this region, curvatures are of order $m^2$. Therefore, although formally we can perform the reduction, the 5d description in terms of the smooth cap is more appropriate.\footnote{Relatedly, varying the asymptotic values of fields between distant points in moduli space causes the set of light states to undergo a large change, leading to a breakdown of effective field theory~\cite{swampland1,swampland2,Heidenreich:2015wga}. Since the excursions that we study are localized, this breakdown does not arise directly, but the need for a 5d description near the bubbles is similar.} The large $\Delta\varphi$ could be interpreted as an artifact of the choice of field space coordinates.  

However, a form of censorship is still at work: although the 5d geometries are smooth, the bubbles are unstable against classical perturbations. For the ES bubble, the unstable mode was first found numerically in the study of black hole nucleation at finite temperature~\cite{grossperryyaffe} and is related by analytic continuation to the Gregory-Laflamme instability of the black string~\cite{gregorylaflamme}. It can be shown that larger bubbles expand~\cite{corleyjacobson}. 

A more intuitive way to understand static bubble instabilities is by analogy. In 4d scalar field theories with metastable vacua, $O(4)$-symmetric bounce solutions mediate vacuum decay. The theories also exhibit static, $O(3)$-symmetric ``sphaleron" solutions. The static bubble sits at the top of the energy barrier under which the bounce solution tunnels. This can be easily visualized in the thin-wall limit, where the tunneling process can be studied in terms of a collective coordinate for the bubble radius $R$ with a potential $V(R)\sim R^2-R^3$ (or similar.) The tunneling bubble nucleates at the finite value of $R$ where $V(R)$ vanishes. In this picture, the ``sphaleron" is the static, massive solution at the value of $R$ where $V(R)$ has its local maximum. It is unstable against perturbations in $R$. 

In KK theory, the ES bubble has a similar relationship to a bounce solution, the semiclassical BON instability of flat KK spacetime~\cite{BON}. The ES bubble may be thought of as the static, massive solution sitting at the top of the energy barrier through which Witten's instanton tunnels to a zero-energy BON~\cite{brillhorowitz,dineshomersun}. It is therefore not surprising that it is classically unstable. We will provide an analogous construction for the EK bubbles, and give a general argument that the ``energy barrier" picture indeed implies instability for the static solution at the local maximum.

Although the ES bubble is a member of the EK bubble family, because the asymptotic behavior is different, it is  convenient to discuss the cases separately. We start with the simpler case of the ES bubble.

\subsection{ES bubble}
\label{sec:ES}
Initial data were constructed in~\cite{brillhorowitz,dineshomersun} exhibiting the instanton--sphaleron relationship for the ES bubble. These data describe spherical, zero-momentum bubbles interpolating from flat KK space, to the static ES bubble, to the Witten bubble, and onward to configurations of arbitrarily negative energy. In fact, the data are simply a 1-parameter family of 4d Euclidean Reissner-Nordstrom geometries,
\begin{align}
ds^2=r^2(d\theta^2+&\sin^2\theta d\phi^2)+(r^2/\Delta)dr^2+(\Delta/r^2) dx_5^2\nonumber\\
\Delta&= r^2-2mr+Q^2\;.
\label{eq:ERN}
\end{align}
These metrics are valid zero-momentum initial data for KK theory because the constraint equations reduce to $\Ricci_4=0$, which is satisfied by~(\ref{eq:ERN}). The $x_5$ periodicity can be chosen so that the spacetime ends smoothly on a bubble of nothing at a finite value of $r$,
\begin{align}
r&\geq r_H=m+\sqrt{m^2-Q^2} \;,\nonumber\\
x_5&\sim x_5+2\pi\gamma\;,\;\;\;\;\;\;\;\;\gamma=\frac{r_H^2}{\sqrt{m^2-Q^2}}\;.
\label{eq:x5perRN}
\end{align}
\begin{figure}[t!]
\begin{center}
\includegraphics[width=0.65\linewidth]{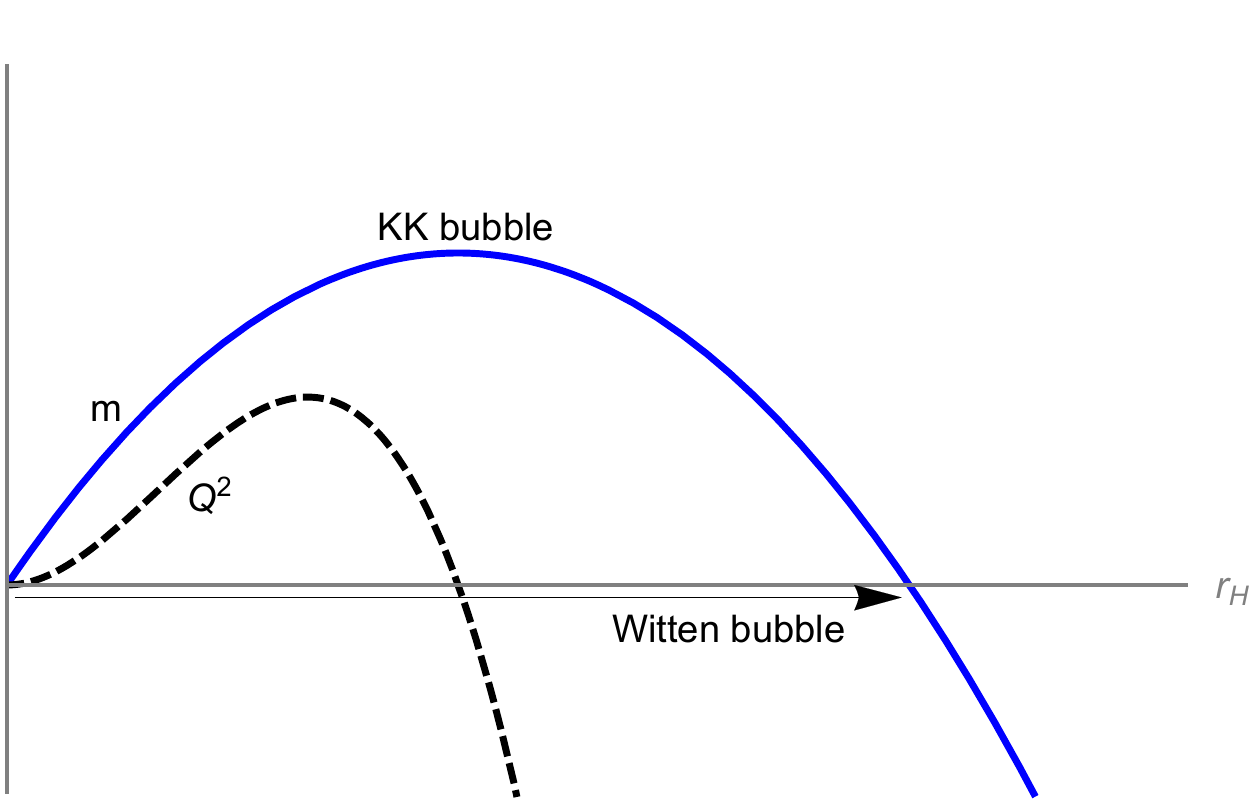}
\caption{
A family of Euclidean Reissner-Nordstrom initial data with bubble radius $r_H$ and fixed asymptotic circle size $\gamma$. The blue solid curve shows the parameter $m$, which is proportional to the energy. The black dashed curve shows the parameter $Q^2$, magnified for clarity. Flat space and the nucleation point of the Witten bubble lie at zero energy, while the static ES bubble sits at a maximum of the energy with $Q=0$. 
} 
\label{fig:ESdata}
\end{center}
\end{figure} 
The relevant 1-parameter family is then determined by fixing the size of the asymptotic KK circle, which is just the $x_5$ periodicity $\gamma$ in Eq.~(\ref{eq:x5perRN}) since $g_{55}\rightarrow 1$ at large distances. In terms of the bubble radius $r_H$ and fixed $\gamma$, 
\begin{align}
m&=r_H-r_H^2/\gamma\nonumber\\
Q^2&=r_H^2-2r_H^3/\gamma\;.
\label{eq:ma2}
\end{align}
The parameters $(m, Q^2)$ of an example family of configurations are shown in Fig.~\ref{fig:ESdata}. For fixed $\gamma$, the energy is proportional to $m$, which has zeros at
\begin{align}
r_H&=0\;\;\;\;\;\;{\rm (flat~KK~space)}\nonumber\\
r_H&=\gamma\;\;\;\;\;\;{\rm (nucleating~BON)}\;.
\end{align}
The latter coincides with the 4d bubble that nucleates after the standard BON tunneling process~\cite{BON}. The static ES bubble appears where $Q^2=0$ and coincides with the maximum of the energy curve.  
The classical instability of the static bubble is a consequence of sitting at the top of the hill, and may be shown manifestly by perturbative analysis~\cite{grossperryyaffe,perryetal}. The solution also controls nucleation of bubbles of nothing at finite temperature~\cite{brownthermal}.

\subsection{EK bubbles}
\label{sec:EK}

The EK bubbles~(\ref{eq:EKline}) are also smoothly connected by a family of zero-momentum initial data to both flat KK space and the nucleation point of a 5d gravitational instanton. The instanton is obtained by analytic continuation of the 5d Kerr metric~\cite{myersperry} and possesses twisted periodicities similar to~(\ref{eq:twisted}). It was constructed in~\cite{dowkeretal}, where it was shown to mediate both BON decays of flat KK space with twisted periodicities and monopole-antimonopole pair creation in the background 4d magnetic field induced by the twists. 

The family of initial data we require is given by a subset of the 4d Euclidean Kerr-Newman (EKN) solutions with zero initial momentum. The EKN line element is given by the spatial part of~(\ref{eq:EKline}) with the replacement
\begin{align}
\Delta= r^2-2m&r-a^2+Q^2\;.
\end{align}
The periodicities required for a smooth geometry are again of the form~(\ref{eq:twisted}), $(x_5,\,\phi)\sim \left(x_5+2\pi\gamma,\,\phi+2\pi \Omega\gamma\right)$, with
\begin{align}
r_H=&m+\sqrt{m^2+a^2-Q^2}\nonumber\\
\gamma=&\frac{r_H^2-a^2}{\sqrt{m^2+a^2-Q^2}}\nonumber\\
\Omega=&\frac{a}{r_H^2-a^2}\;.
\end{align}
For fixed $\gamma$, the energy of these configurations is proportional to $m$ (see Section \ref{sec:generalinstab}).

\begin{figure}[t!]
\begin{center}
\includegraphics[width=0.65\linewidth]{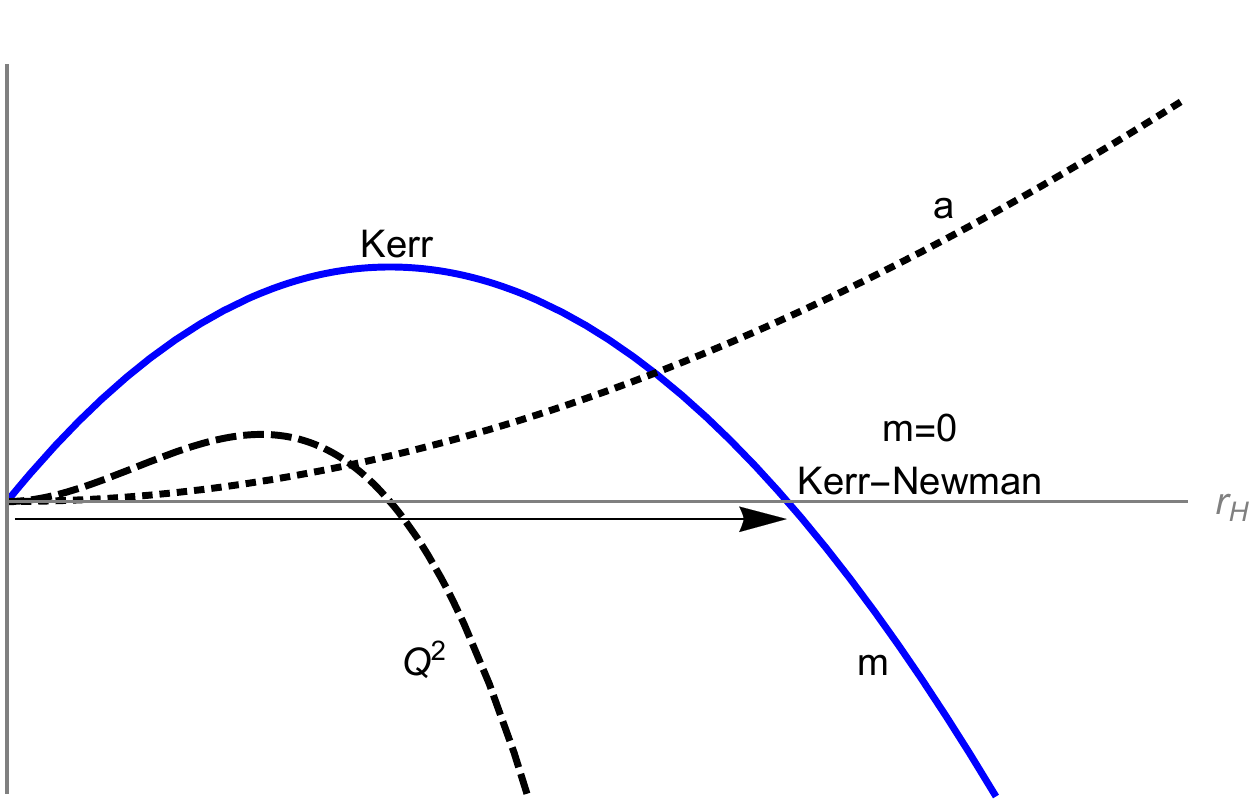}
\caption{A family of Euclidean Kerr-Newman initial data with bubble radius $r_H$ and representative values for the fixed asymptotic periodicities $(\Omega,\gamma)$. The blue solid curve shows the parameter $m$, which is proportional to the energy. The black dashed (dotted) curve shows the parameter $Q^2$ (a), magnified for clarity. Flat space and the nucleation point of the instanton studied in~\cite{dowkeretal} lie at zero energy, while the static EK bubble sits at a maximum of the energy with $Q=0$. 
} 
\label{fig:EKNdata}
\end{center}
\end{figure}

As in the previous case, we obtain a 1-parameter family by holding the asymptotic behavior of the initial data fixed. This amounts to two constraints on the parameters $(m,a,Q^2)$ given by $\gamma=const$, $\Omega=const$. In terms of the EKN bubble radius $r_H$, we obtain 
\begin{align}
m&=\frac{1+2\gamma\Omega^2 r_H-\sqrt{1+4 r_H^2\Omega^2}}{2\gamma\Omega^2}\nonumber\\
a&=\frac{-1+\sqrt{1+4r_H^2\Omega^2}}{2\Omega}\nonumber\\
Q^2&=\frac{\gamma +4 \gamma  r_H^2 \Omega^2+2
   r_H-(\gamma+2r_H)  \sqrt{4 r_H^2
   \Omega^2+1}}{2 \gamma  \Omega^2}\;.
\label{eq:maQ2}
\end{align}
The static bubbles have $Q=0$, which implies $|\gamma\Omega|<1$. Therefore the families for which $|\gamma\Omega|\geq1$ miss the static solutions as $r_H$ is varied. If $|\gamma\Omega|<1$, there is a static solution in the family and the large-$r_H$ behavior of $m$ is negative. The parameters $(m,a,Q^2)$ of such an example family of configurations are shown in Fig.~\ref{fig:EKNdata}.

The parameter $m$ has two zeros as a function of $r_H$,
\begin{align}
r_H&=0\;\;\;\;\;\;\;\;\;\;\;\;\;\;\;\;\;\;\;\;\;{\rm (flat~KK~space)}\nonumber\\
r_H&=\frac{\gamma}{1-(\gamma\Omega)^2}\;\;\;\;\;\;{\rm (nucleating~BON)}\;.
\end{align}
The second point matches the bubble radius that nucleates at the end of a BON tunneling event~\cite{dowkeretal}. Furthermore, it is easily checked that the maximum value of $m$ in Eq.~(\ref{eq:maQ2}) occurs at a zero of $Q^2$. Thus the peak coincides with the static EK bubble, as expected.

To summarize, the ES/EK bubble spacetimes provide smooth 5d UV completions of divergent localized scalar excursions in 4d. However, the forces that cancel to produce a static bubble  leave it in unstable equilibrium. The bubbles either collapse into black strings or catastrophically expand.

It is interesting to note that wound black string solutions in KK space are also unstable when the radion approaches the string scale outside the horizon~\cite{Horowitz:2005vp}. In this case tachyon condensation of closed strings on the black string background signals the instability, and the endpoint is a bubble of nothing~\cite{Horowitz:2005vp}. These solutions thus provide a similar class of examples to the KK bubbles, albeit ones in which the instability involves intrinsically stringy dynamics.

\section{Instabilities and $\delta^2 E<0$: general analysis}
\label{sec:generalinstab}

\noindent In this section we will argue that the property established above for the static KK bubbles, that they locally maximize the energy along certain one-parameter families of initial data, implies that they possess a classical instability. Although such instability is intuitively plausible by analogy with simple mechanical systems, further analysis is required to establish the result in general relativity. Our discussion will be somewhat technical, but not completely rigorous; in one step, for example, we will assume a symmetric operator has a self-adjoint extension without proof. We will indicate where such assumptions are made. 

In brief, we will show that metric perturbations $\delta g$ around static vacuum solutions satisfy an equation of the form $\ddot{\delta g}\sim -S[\delta g]$, where $S$ is a linear differential operator, and for a certain class of perturbations, $\int \delta g S[\delta g]$ is the second order perturbation of the energy. $S$ is a symmetric operator with domain given by square-integrable perturbations satisfying the constraint equations, and $S$ can be shown to annihilate residual infinitesimal gauge transformations. Therefore, if a perturbation in the class lowers the energy, $S$ must have a negative eigenvalue. This mode grows exponentially in time, and the instability is physical, since it cannot be removed by a gauge transformation. Finally, we show that the families of initial data identified for the KK bubbles generate the relevant type of perturbations around the static solutions, from which instability follows.

In vacuum, the canonical variables (spatial metric $g_{ab}$ and conjugate momentum $\pi^{ab}$) satisfy the following equations:
\begin{eqnarray}
\dot{g}_{ab} &=& 2\dfrac{N}{\sqrt{g}}\left(\pi_{ab}-\dfrac{1}{D-1}g_{ab}\,\pi\right)+2\nabla_{(a}N_{b)},\label{hamg}\\
\dot{\pi}^{ab} &=& -N\sqrt{g}\,G^{ab} + Q^{ab},\label{hamp}
\end{eqnarray}
where $N$ and $N_a$ are the lapse and the shift functions, and $D>1$ is the spatial dimension. All the metric-related quantities -- the Ricci and the Einstein tensor $R_{ab}$ and $G_{ab}=R_{ab}-1/2\,R\,g_{ab}$, the covariant derivative $\nabla_a$, and the determinant $g$ in the given coordinates -- are associated with the spatial metric, which is also used for lowering and raising indices. The form of $Q_{ab}$ can be found in~\cite{Wald:1984rg}. We do not write it explicitly because when we fix the gauge below it will not contribute to the linearized equation. The Hamiltonian and momentum constraints are:
\begin{eqnarray}
\mathcal{H}&=&\left(\pi^{ab}\pi_{ab}-\dfrac{1}{D-1}\pi^2\right) - \sqrt{g}\,R=0,\label{hamconstr}\\
\mathcal{H}_a&=&\sqrt{g}\,\nabla_b\dfrac{\pi^{ab}}{\sqrt{g}}=0.\label{momconstr}
\end{eqnarray}
Equation \eqref{hamg} can be solved for $\pi^{ab}$:
\begin{equation}
\label{mom}
\pi^{ab}=\dfrac{\sqrt{g}}{2N}K^{abcd}(\dot{g}_{cd} - 2\nabla_{\!(c}N_{d)}),
\end{equation}
where
\[
K^{abcd}=\dfrac{1}{2}\big(g^{ac}g^{bd}+g^{ad}g^{bc}\big)-g^{ab}g^{cd}.
\]
Consider a one-parameter family of vacuum solutions that reduces to a static solution at $\lambda=0$ for which the lapse function is $1$ and the shifts are zero:
\begin{equation}
\stat{R}_{ab}=0,\;\;\;\stat{\pi}^{ab}=0,\;\;\;\stat{N}=1,\;\;\;\stat{N}_a=0.
\label{stat_sol}
\end{equation}
The hat indicates the static solution and any quantity associated with it. A (first order) perturbation is defined by $\delta f=\partial_\lambda f|_{\lambda=0}$. We consider perturbations only about the static solution, so whenever we say ``perturbation,'' it is always meant about the static solution. The static metric is used to raise and lower the indices of a perturbation of a tensor. We impose the gauge condition 
\begin{equation}
\label{gauge}
N=1,\;\;\;\;\;N_a=0.
\end{equation}
The static spacetime metric is given in Gaussian normal coordinates ($\stat{N}=1$ and $\stat{N}_a=0$), and the gauge condition means that the perturbed spacetime metric is also written in such coordinates.  This choice simplifies the analysis. To get the linearized (perturbed) equation satisfied by $\delta g_{ab}$, we differentiate Eq.\ \eqref{hamp} with respect to $\lambda$ at $\lambda=0$ and use Eq.\ \eqref{mom}. $Q^{ab}$ is the sum of a term linear in $\partial_aN$ and a quadratic expression of $\pi_{ab}$ and $N_a$, so by Eq.\ \eqref{gauge} and $\hat{\pi}^{ab}$, it does not contribute to the linearized equation. We obtain 
\begin{equation}
\dfrac{\sqrt{\stat{g}}}{2}\,\hat K^{abcd}\,\delta\ddot{g}_{cd}=-\delta(\!\sqrt{g}\,G^{ab})=-\sqrt{\stat{g}}\,S[\delta g]^{ab},
\label{lin_dyn}
\end{equation}
where we introduced the linear operator $S$ acting on $\delta g_{ab}$. 

We now show that $S$ is related to the energy, and that it is a symmetric operator in a suitable scalar product, which will allow us to study the time evolution of its eigenmodes. We will need the $\lambda$-derivative of the Ricci scalar, 
\begin{equation}
\label{pert_ricci}
\partial_\lambda(\!\sqrt{g} R) = \sqrt{g}\,[\nabla^a(\nabla^b\partial_\lambda g_{ab}-g^{cd}\nabla_a\partial_\lambda g_{cd})-G^{ab}\partial_\lambda g_{ab}].
\end{equation}

Take a two-parameter family of  solutions. The parameters are $\lambda_1$ and $\lambda_2$, and the corresponding perturbations are $\delta_1g_{ab}$ and $\delta_2g_{ab}$. The static solution is at $\lambda_1=\lambda_2=0$. If we write  Eq.\ \eqref{lin_dyn} for $\delta_1g_{ab}$ and contract it with $\delta_2g_{ab}$, the right hand side is $-\delta_2g_{ab}\,\delta_1(\!\sqrt{g}\,G^{ab})$, which using Eq.\ \eqref{pert_ricci} is equal to the following expression evaluated at $\lambda_1=\lambda_2=0$:  
\begin{align}
\sqrt{g}\,G^{ab}\partial_{\lambda_1}\partial_{\lambda_2}g_{ab}-\partial_{\lambda_1}(\!\sqrt{g}\,G^{ab}\partial_{\lambda_2}g_{ab})
=\,&\sqrt{g}\,G^{ab}\partial_{\lambda_1}\partial_{\lambda_2}g_{ab}+\partial_{\lambda_1}\partial_{\lambda_2}(\!\sqrt{g}\,R) \nonumber\\
&-\partial_{\lambda_1}\big[\sqrt{g}\,\nabla^a(\nabla^b\partial_{\lambda_2} g_{ab}-g^{cd}\nabla_a\partial_{\lambda_2} g_{cd})\big].
\label{eq:Scontract}
\end{align}
The first two terms on the right-hand side are manifestly symmetric. So if it was not for the last term, the spatial integral of $\delta_2g_{ab}\,\delta_1(\!\sqrt{g}\,G^{ab})$ would take the desired form $\langle\delta_2 g, S\delta_1 g\rangle$ with a symmetric operator $S$ and the usual $L^2$ scalar product of symmetric tensors $s_{ab}$ and $t^{ab}$,
\begin{equation}
\label{innerprod}
\langle t,s\rangle=\int\d^D\!x\sqrt{\stat{g}}\,t^{ab}s_{ab}.
\end{equation}

However, requiring asymptotic flatness (discussed further below), the integral of the last term in Eq.\ \eqref{eq:Scontract} is in fact symmetric: 
\begin{align}
\int\d^D\!x\,\partial_{\lambda_1}\big[\sqrt{g}\,\nabla^a&(\nabla^b\partial_{\lambda_2} g_{ab}-g^{cd}\nabla_a\partial_{\lambda_2} g_{cd})\big]\nonumber\\
&=\lim\partial_{\lambda_1}\oint\,\d^{D-1}\!x\,\sqrt{h}\,n^a(\nabla^b\partial_{\lambda_2} g_{ab}-g^{cd}\nabla_a\partial_{\lambda_2} g_{cd})\nonumber\\
&=\lim\oint\,\d^{D-1}\!x\sqrt{h}\,n^a(\nabla^b\partial_{\lambda_1}\partial_{\lambda_2} g_{ab}-g^{cd}\nabla_a\partial_{\lambda_1}\partial_{\lambda_2} g_{cd}).
\label{eq:ssymm}
\end{align}
Here the limit is taken over an increasing sequence of compact spatial regions that cover the entire space, $\oint$ is the integral on the boundary of these regions, $n^a$ is the outward unit normal to these boundaries, and $h$ is the determinant of the induced metric. The last equality holds because there is no contribution from the terms generated by $\partial_{\lambda_1}$ acting on metric components in $n^a$, $h$, or the Christoffel symbols of the covariant derivatives. If the first order variation preserves asymptotic flatness (discussed further below), these terms decay more rapidly than the surface area of the boundary. In this calculation, we repeatedly changed the order of differentiations, limits, and integrations. Since we are not aiming for mathematical rigor, the validity of these steps will not be checked here. 

We can get $\langle\delta g, S\delta g\rangle$ from the second derivatives along a single family by essentially the same calculation.  To summarize, we have 
\begin{align}
\langle\delta_2 g, S\delta_1 g\rangle&=\bigg[ -\int\d^D\!x\, \partial_{\lambda_1}\partial_{\lambda_2}(\!\sqrt{g}\,R) \nonumber\\
&~~~~~~~ + \lim\oint\,\d^{D-1}\!x\sqrt{h}\,n^a(\nabla^b\partial_{\lambda_1}\partial_{\lambda_2} g_{ab}-g^{cd}\nabla_a\partial_{\lambda_1}\partial_{\lambda_2} g_{cd})\bigg]\bigg|_{\lambda_1=\lambda_2=0},\\
 \langle\delta g, S\delta g\rangle&= \bigg[-\int\d^D\!x \,\partial^2_{\lambda}(\!\sqrt{g}\,R) +\lim\oint\,\d^{D-1}\!x\sqrt{h}\,n^a(\nabla^b\partial_\lambda^2g_{ab}-g^{cd}\nabla_a\partial_\lambda^2g_{cd})\bigg]\bigg|_{\lambda=0}.
\label{adm}
\end{align}
Here we have used that $\hat G_{ab}=0$ by Eq.\ \eqref{stat_sol}.

For an asymptotically flat spacetime, there are coordinates in which the metric components approach the components of the $D+1$ dimensional Minkowski metric in Cartesian coordinates at least as fast as $r^{2-d}$, where $r$ is the Euclidean length calculated from the spatial coordinates $x^i$ and $d\geq 3$ is the number of noncompact dimensions. Furthermore, every differentiation with respect to a noncompact coordinate contributes an additional factor of $r^{-1}$ to the asymptotic scaling. In such coordinates, the (total) energy of an asymptotically flat spacetime is
\begin{equation}
\label{total_energy}
E=\frac{1}{16\pi G_{D+1}}\lim_{r\to\infty}\oint\,\d^{D-1}\!x\sqrt{h_0}\,\sum_j\frac{x^i}{r}(\partial_j g_{ij}-\partial_i g_{jj}),
\end{equation}
where $G_{D+1}$ is the $D+1$ dimensional Newton constant, and $d^{D-1}\!x\sqrt{h_0}$ is the invariant measure induced by the $D$ dimensional Euclidean metric on the boundary surface of constant $r$, over which the integral $\oint$ is taken. Repeating the argument by which the dropping of additional terms in the last line of Eq.\ \eqref{eq:ssymm} is justified, one finds that
\begin{equation}
\label{2ndvarE}
\langle\delta g, S\delta g\rangle= 16\pi G_{D+1}\,\delta^2E-\int\d^D\!x\sqrt{\hat g}\,\delta^2R\;,
\end{equation}
where $\delta^2$ is the second derivative with respect to $\lambda$ at $\lambda=0$.

Asymptotically flat metrics decay rapidly enough for the energy to be finite, but not too rapidly, so that the energy is typically not zero. The same asymptotic condition on perturbations in the domain of $S$ would  not necessarily imply square-integrability for $d\leq 4$. However, Eq.\ \eqref{adm} already indicates that perturbations satisfy stronger asymptotic conditions. The boundary term depends only on the second order perturbation $\partial_\lambda^2 g_{ab}|_{\lambda=0}$, while the other terms depend only on first order perturbations. (That this is the case for the $\partial_\lambda^2(\sqrt{g}R)$ term follows from the Hamiltonian constraint~(\ref{hamconstr}) and $\hat\pi^{ab}=0$. Then this term is quadratic in the first order momentum perturbation $\partial_\lambda\pi^{ab}$.) But a one-parameter family $\lambda\mapsto g_{ab}(\lambda)$ of metrics can always be reparametrized as $\lambda\mapsto g_{ab}(\lambda^2/2)$. The first order perturbations are zero in the new parametrization, and the second order perturbation is the same as the first order perturbation in the old parametrization. Eq.\ \eqref{adm} holds in any parametrization, and only the boundary term survives in the new parametrization, where it becomes the first order perturbation $\delta E$ of the energy in the original parametrization. Therefore, we conclude that $\delta E=0$ on a static background. This is analogous to the principle of virtual work in mechanics. It is also a special case of the first law of thermodynamics of gravitational physics, which takes the simple form of $\delta E=0$ if both the entropy and the terms related to the change of the angular momentum or the electromagnetic charge are zero (for a detailed discussion, see~\cite{Hollands:2012sf}).

Now suppose that the families consist of metrics that admit (in coordinates used in our characterization of asymptotically flat spacetimes) a power series expansion in terms of $r^{-1}$, and the subleading term, which behaves as $r^{2-d}$, is proportional to the energy. Then $\delta E=0$ implies that the perturbation decays at least as rapidly as $r^{1-d}$, which is also fast enough for square-integrability. We will consider only perturbations that have this property.

Another restriction on the physically relevant perturbations comes from the constraints. By equations \eqref{stat_sol} and \eqref{pert_ricci}, the perturbation of the Hamiltonian constraint \eqref{hamconstr} gives
\begin{equation}
\label{lin_hamconstr}
\partial_\lambda(\!\sqrt{g}R)|_{\lambda=0}=\sqrt{\stat{g}}\big(\stat{\nabla}^a\stat{\nabla}^b-\stat{g}^{ab}\stat{\nabla}^c\stat{\nabla}_{\!c}\big)\,\delta g_{ab}=0.
\end{equation}
This condition is invariant under $S$. Actually, $(S\delta g)^{ab}$ satisfies an even stronger condition if the above equation holds for $\delta g_{ab}$:
\begin{eqnarray}
\stat{\nabla}_a(S\delta g)^{ab}&=&\stat{\nabla}_a\sqrt{\stat{g}}^{-1}\delta(\!\sqrt{g}\,G^{ab})=\partial_\lambda(\nabla_aG^{ab})|_{\lambda=0}=0,\label{transverse}\\
\stat{g}_{ab}(S\delta g)^{ab}&=&\frac{2-D}{2}\partial_\lambda R|_{\lambda=0}=0.\label{traceless}
\end{eqnarray}
Eqs.\ \eqref{transverse} and\ \eqref{traceless} are obtained by replacing $\stat{g}_{ab}$ by the $\lambda$ dependent metric $g_{ab}$ and moving all terms in front of $\partial_\lambda$. This step is permissible because when $\partial_\lambda$ acts on a metric component that was originally a component of $\stat{g}_{ab}$, it produces a linear expression in $\stat{G}^{ab}$ or $\stat{R}$, both of which are zero. Once all terms are inside the $\lambda$ differentiation, we can use the transversality of the Einstein tensor (true for any metric) and $\delta R=0$ (true for any perturbation of a static solution) to obtain Eqs.\ \eqref{transverse} and~\eqref{traceless}. Therefore $(S\delta g)^{ab}$ is transverse and traceless, from which $(\stat{\nabla}_a\stat{\nabla}_b-\stat{g}_{ab}\stat{\nabla}^c\stat{\nabla}_{\!c})\,(S\delta g)^{ab}=0$ immediately follows.

Now consider gauge transformations, generated by a vector field $\xi^a$ parametrized as 
\begin{equation}
\label{xi_param}
\xi^0=\dfrac{\alpha}{N},\;\;\;\xi^i=\beta^i-\dfrac{\alpha}{N}N^i.
\end{equation}
The infinitesimal gauge transformation of the lapse and shift functions generated by $\xi^a$ is
\begin{equation}
\label{n_inf}
\begin{aligned}
\delta_\xi N&=\partial_0\alpha+ \beta^i\partial_iN -N^i\partial_i\alpha,\\
\delta_\xi N^i&=\partial_0\beta^i+g^{ij}(\alpha\partial_jN-N\partial_j\alpha)+\beta^j\partial_jN^i-N^j\partial_j\beta^i.
\end{aligned}
\end{equation}
Infinitesimal gauge transformations of the static solution are also perturbations: they are obtained when the family of metrics is generated by a one-parameter family of diffeomorphisms applied to the static solution. In this case $N=\stat{N}$ and $N^i=\stat{N}^i$ on the right-hand sides of Eq.\ \eqref{xi_param} and \eqref{n_inf}, so the gauge condition $\delta N=0$ and $\delta N_a=0$ is preserved if and only if $\alpha$ is time independent and $\beta^i=x^0\hat g^{ij}\partial_j\alpha+\gamma^i$, where $\gamma^i$ is also time independent. The second time derivative of a transformation $\delta_\xi\stat{g}_{ab}$ generated by such a vector field is zero, so they are in the kernel of $S$ by Eq.\ \eqref{lin_dyn}. This can also be checked by direct calculation using Eq.\ \eqref{pert_ricci} and the specific form of $\delta g_{ab}$.

We have shown that $S$ is a symmetric operator with respect to the inner product \eqref{innerprod}. The domain $\mathcal{D}$ of $S$ is the linear space of perturbations decaying at least as fast as $r^{1-d}$, so they are square integrable. They also satisfy Eq.\ \eqref{lin_hamconstr}. This condition is invariant under $S$ and $S[\delta g]^{ab}$ decays even faster than $r^{1-d}$, so $S[\mathcal{D}]\subset\mathcal{D}$. Therefore $S$ can be thought of as an operator on the Hilbert space obtained by completing $\mathcal{D}$ with respect to the norm $\|\delta g\|^2=\langle\delta g,\delta g\rangle$. Although $S$ is only symmetric, we assume that it is essentially self-adjoint. By the spectral theorem applied to its self-adjoint extension $\bar{S}$, any element of $\mathcal{D}$ is decomposed into eigenvectors of $\bar{S}$ (where the decomposition may involve integration if the spectrum of $\bar{S}$ has a continuous part). Although these eigenvectors may not be as smooth as the original perturbations in $\mathcal{D}$, we will imagine that they are. Since we do not elaborate on technical issues hinted at in this paragraph, our analysis remains somewhat heuristic.

Let $\gamma_{ab}$ be an eigenvector of $S$ with eigenvalue $\kappa$. Since $\gamma_{ab}$ is an eigenvector, it satisfies not only Eq.\ \eqref{lin_hamconstr}, but also the stronger conditions \eqref{transverse} and \eqref{traceless} satisfied by $S[\gamma]^{ab}$. In particular, it is traceless, so $K^{abcd}\gamma_{cd}=\gamma^{ab}$ and 
\begin{align}
\delta g_{ab}=\gamma_{ab}\exp(\pm\sqrt{-2\kappa}\,t)
\end{align}
 solves equation \eqref{lin_dyn}. It also solves Eq.\ \eqref{lin_hamconstr}. The perturbation of the momentum constraint \eqref{momconstr} is
\[
\stat{\nabla}_a\dfrac{\delta\pi^{ab}}{\sqrt{\stat{g}}}=0,
\]
which is also satisfied because $\delta\pi_{ab}=\sqrt{\stat{g}}\,\delta\dot{g}_{ab}$ by the tracelessness of $\gamma_{ab}$ and $\delta N_a=0$, and $\gamma_{ab}$ is also transverse. If $\kappa<0$, one of the solutions is exponentially growing in the future. This behavior cannot be due to the choice of coordinates because none of the infinitesimal gauge transformations allowed by Eq.\ \eqref{gauge} change $\delta\ddot{g}_{ab}$, so adding such a transformation to our solution will not remove the exponential growth. Therefore, a negative eigenvalue of $\bar{S}$ indicates physical instabilities of the static solution.

If we have a family along which the second variation $\delta^2E$ of the energy is negative and $\delta^2 R=0$, then by Eq.\ \eqref{2ndvarE}, the decomposition of the corresponding first order perturbation into eigenvectors of $\bar{S}$ must have an element with a negative eigenvalue, so the static solution is unstable. This is analogous to the mechanical instability at a local maximum of the potential. (By the Hamiltonian constraint, the vanishing of $\delta^2 R$ is analogous to probing the potential without generating kinetic energy.) It is also a special case of thermodynamic instability in gravitation, whose condition reduces to the form $\delta^2 E<0$ if the entropy and the work terms associated with the angular momentum and the electromagnetic charge are zero~\cite{Hollands:2012sf}.

The total energy \eqref{total_energy} of an EK bubble is $E=\pi\gamma m/G_5$. Thus $\delta m=0$ at the static solution in a family of bubble initial data (parametrized by $\lambda=r_H$) with fixed $(\Omega,\gamma)$. In fact, as can be seen from Eq.\ \eqref{eq:maQ2}, the static bubble sits at a maximum of $m$, so $\delta^2E<0$. The subleading term in the $1/r$ expansion of the spatial metric is proportional to $m/r$, so $\delta m=0$ implies that $\delta g_{ab}$ decays as $1/r^2$ and therefore it is square-integrable. Furthermore, $R=0$ for each metric in the family. So the perturbations have all the properties needed for our analysis, and we can conclude that the static EK bubbles are unstable.

\section{Strings on the ES bubble background}
\label{sec:string}
It is also interesting to ask what sorts of couplings to other degrees of freedom are diagnostic of the local value of the KK scalar. (This question is, however, somewhat tangential to the previous sections.)  Closed strings are a natural probe: strings of fixed tension $T$ wound around the compact direction acquire mass sensitive to the local value of $\varphi$. In the asymptotic region the wound strings are stretched and heavy, with masses $\sim mnT$.\footnote{Spacetime fermions must have antiperiodic boundary  conditions on BON backgrounds. On ordinary KK backgrounds, antiperiodic BCs lead to tachyonic string modes in sectors of odd winding number if the KK radius is smaller than the string scale and slowly varying in space~\cite{Rohm:1983aq}, which can catalyze bubble production~\cite{Adams:2005rb,Horowitz:2005vp}.}

Although the static ES bubble is unstable, we can imagine finely-tuned, quasi-static initial conditions near the energy peak, and introduce closed strings onto this background.  Near the core of the bubble, the size of the extra dimension goes to zero in a smooth cap. Physically, we expect that the wound string can shrink and slip off the cap, converting itself to outgoing unwound states of lower mass $\sim\sqrt{T}$.

This intuitive picture can be supported by analyzing a simplified system, retaining only a subset of the wound string degrees of freedom.  
We consider a 4d field $\Phi$ with action
\begin{align}
S&=\int d^4x \sqrt{-G} \left(-\frac{1}{2}G^{\mu\nu}\partial_\mu \Phi \partial_\nu \Phi -\frac{1}{2}(8\pi m n T)^2V\Phi^2\right)\nonumber\\
&=\int d^4x \sqrt{-g} \left(-\frac{1}{2}\frac{g^{\mu\nu}\partial_\mu \Phi \partial_\nu \Phi}{\sigma^{1/3}} -\frac{1}{2}(8\pi m n T)^2\Phi^2\right)
\label{eq:stringL}
\end{align}
where in the second line we have made the conformal transformation to Einstein frame, and the metric and KK scalar were defined in Eqs.~(\ref{eq:metparam}) and~(\ref{eq:ESfourd}). This is a toy model for wound strings on the ES bubble background; $\Phi$ creates particle states corresponding microscopically to strings of winding number $n$, with no other excitations. 

(We can neglect the additional terms in the string energy of order $\sqrt{T}$ by restricting to large winding numbers $n$, but in any case we are mainly interested in seeing whether energies $\sim\sqrt{T}$ can emerge given mass-squared terms $\sim T^2$, as in~(\ref{eq:stringL}). We also note that a related analysis was performed in~\cite{dineshomersun}, with somewhat different results.)

In the complete system, wound strings should slip off the cap of the extra dimension.  This process cannot be described with $\Phi$ alone due the symmetry of the simplified system, but since $\Phi$ becomes light near $r=2m$,  we can  look instead for bound states. Sufficiently deeply bound states of the simplified system with energies of order $\sqrt{T}$ indicate that real strings can shed their winding number completely by being thrown into the bubble.

The Klein-Gordon equation arising from Eq.~(\ref{eq:stringL}) is
\begin{align}
\label{eq:KG-gen}
(\Box_G -m_X^2)\Phi = 0\;,
\end{align}
where
\begin{align}
m_X^2=m_0^2 V=m_0^2 f&,\;\;\;\;\;
m_0\equiv8\pi mnT\;. 
\end{align}
We look for $\ell=0$ solutions of the form $\Phi=X(r)e^{i\omega t}$. 
Expanding Eq.~\eqref{eq:KG-gen},
\begin{align}
f\,X''(r)+\left(\frac{1+3f}{2r}\right)
   X'(r)- m_0^2\,f\,X(r)=-\omega^2X(r)\;,
    \label{eq:KG}
 \end{align}
In normal form, $X=u/\left(rf^{1/4}\right)$, Eq.~(\ref{eq:KG}) becomes
\begin{align}
-f u'' + f W u = \omega^2 u\;,
\label{eq:normal}
\end{align}
where
\begin{align}
\label{potential}
W=-\frac{3}{4}\frac{m^2}{f^2r^4}+m_0^2\;.
\end{align}
We will not attempt to solve Eq.~(\ref{eq:normal}) exactly. It is much simpler to place bounds on the ground state energy and see that the expected scaling emerges.

We begin by establishing a simple upper bound on the ground state energy which exhibits $\sqrt{T}$ scaling. We will need the conserved Klein-Gordon scalar product, which on $\ell=0$ modes is given by the radial integral
\begin{align}
I[u^*v]\equiv \int_{2m}^\infty dr\, u^*v/f\;.
\label{eq:kgp}
\end{align}
This product differs from the ordinary $L^2$ product by a weight function $f^{-1}$. It may be obtained from the definition of the Klein-Gordon product, or more simply by inspection of the eigenvalue problem \eqref{eq:normal}. The potential would be hermitian with any weight  function $w$ in the integral defining the scalar product, but the term $f u''$ requires $w = f^{-1}$, leading to (\ref{eq:kgp}). The ground state ``energy" $E_0\equiv\omega^2$ has an upper bound of
\begin{align}
E_0\leq \frac{I[u^*(-u'' + W u)] }{ I[ |u|^2 / f ]}\;,
\label{eq:sz1}
\end{align}
where $u$ can be any smooth function satisfying appropriate boundary conditions at $r=2m$. Let $\varphi$ be a nonzero smooth function on $\mathbb{R}^+$ supported within $(0,1)$ and define $u_0(r)=\varphi(m_0(r-2m))$. Since $u_0(r)=0$ for $r-2m>1/m_0$, we have $I[ |u_0|^2 / f ]>2mm_0\,I[|u_0|^2]$, so
\begin{align}
E_0<\frac{I[u_0^*(-u_0'' + W u_0)] }{ 2mm_0\, I[ |u_0|^2 ]}<\frac{I[u_0^*(-u_0'' + m_0^2\,u_0)] }{ 2mm_0\, I[ |u_0|^2 ]}=\frac{m_0}{2m}\frac{\int_{\mathbb{R}^+}\varphi^*(-\varphi'' + \varphi) }{\int_{\mathbb{R}^+} |\varphi|^2 }\sim nT.
\end{align}
We thus obtain an upper bound on the lowest frequency that scales as $\omega\lesssim \sqrt{nT}$. This can be contrasted with the mass of the string $\sim mnT$ in the  asymptotically flat region of the spacetime; the bound modes are parametrically light. We interpret these light modes of the simplified system as reflecting the ability of wound strings to ``slip off" the cigar: wound string states can be converted into unwound states  by throwing them into the region where the KK scalar becomes large. 

Note that $u_0$ is supported away from $r=2m$, so the upper bound we obtained is independent of the precise boundary conditions at $r=2m$. To rigorously establish the scaling of the ground state energies, we need lower bounds on the ground state $\omega$ that also scale as $\sqrt{nT}$ for large $n$. It is easiest to proceed by making a coordinate transformation that removes the singularity at $r=2m$. In these coordinates the question of boundary conditions also becomes more transparent. We give a detailed analysis in Appendix~\ref{appx:lowerbound}; here we only show that the properties of the attractive $1/x^2$ potential in nonrelativistic quantum mechanics suggest that the  $\omega^2$ spectrum is at least positive. 

Since $W(r)> m_0^2 -3/16x^2 > -3/16 x^2$ for $x=r-2m>0$,
\begin{align}
E_0\geq\inf\frac{I[u^*(-u'' + W u)] }{ I[ |u|^2 / f ]}\geq\inf\frac{\int dx\,u^*(-u'' -\frac{3}{16x^2}\,u) }{\int dx\,(1+\frac{2m}{x})|u|^2 }\;,
\end{align}
where the infimum is taken over smooth functions $u$ satisfying appropriate boundary conditions at $x=0$. Thus, if the spectrum of a nonrelativistic particle in the potential $-3/16x^2$ is positive, $E_0\geq 0$ follows. This is a somewhat subtle problem in ordinary quantum mechanics: the potential $-\lambda/x^2$ leads to different behaviors depending on the strength of the coupling $\lambda$. For weak attractive coupling, $0<\lambda\leq 1/4$, it is possible to proceed in the ordinary way, identifying regular and irregular solutions to the Schr\"odinger equation near the origin and connecting the regular solutions with controlled behavior at large $r$. One finds a continuum of scattering states of positive energy and no bound states. For strong attractive coupling, $\lambda>1/4$, it is no longer possible to label the small-$r$ solutions as regular and irregular: both behave as $\sin\log\, r$ and admit an infinite number of nodes below any fixed $r_0$. The spectrum is either unbounded from below, or additional physical input is needed to establish a ground state (see, e.g.,~\cite{schwartz,holstein}.) Since $\lambda=3/16$ is close to but safely below the critical coupling of $1/4$ in the nonrelativistic quantum mechanical problem, we conclude that $E_0\geq 0$.

\section{Discussion}
\label{sec:disc}
In 4d gravity with a light scalar, attempts to set up a large variation of the scalar can result in collapse to a black hole, even if  energy densities are kept small. However, vacuum solutions of Kaluza-Klein theory are known in which the KK scalar diverges in local regions. These regions have a different fate: they are unstable bubbles of nothing which may either collapse or expand. The instability is well-known for the bubble derived from the Euclidean Schwarzschild solution, and we have shown that it is also present for Euclidean Kerr bubbles, which we find can be interpreted as static relatives of bubble nucleation events in KK magnetic fields. An exterior observer can in principle detect the field excursion, for example by throwing a wound string into an approximate KK bubble geometry and watching it emerge as states of zero winding number. However, realizing an exactly static bubble is impossible, and for a finely-tuned quasistatic bubble, in general this experiment will perturb it along the unstable direction. 

These observations add to the collection of explicit examples in which general relativity dynamically censors large scalar field excursions. It would be interesting to explore further whether there is any sharp relationship with more existential conjectures about the admissible types of scalar potentials. 

\vskip 1cm
\noindent
{\bf Acknowledgements:}  PD thanks L. Sorbo, J. Traschen, D. Kastor, M. Dine, T. Banks, G. Horowitz, M. Reece, J. Kozaczuk, and A. Brown. This work was supported by NSF grant PHY-1719642.

\appendix
\section{Lower Bound on $\omega$}
\label{appx:lowerbound}
The transformation $\rho(r)$ that brings the spatial metric into a function times the Euclidean metric satisfies:
\begin{align}
\frac{d\rho}{dr}=\frac{\rho}{r}f^{-1/2}.
\end{align}
Choosing the solution for which $\rho(r_0)=r_0$, where $r_0=2m$, we find
\begin{align}
\rho=2r-r_0+2\sqrt{r(r-r_0)}\;,
\end{align}
and the metric in the new coordinates is the Euclidean metric times $\Omega^2(\rho)=(\rho + r_0)^4/16\rho^4$. In the new coordinates, the normal form of Eq.~(\ref{eq:KG}) is obtained by the substitution $X(r(\rho))=v(\rho)/(\rho\sqrt{\Omega(\rho)})$:
\begin{align}
-\frac{16\rho^4}{(\rho+r_0)^4}v''+m_0^2\left(\frac{\rho-r_0}{\rho+r_0}\right)^2v=\omega^2v
\label{eq:newcoord}
\end{align}
The Klein-Gordon product on radial functions $v(\rho)$ takes the form
\begin{align}
I[v_1^*v_2]=\int_{r_0}^\infty d\rho \left(\frac{\rho+r_0}{2\rho}\right)^4v_1^*v_2\;.
\end{align}
The ground state frequency $\omega^2$ is given by
\begin{align}
r_0^2\omega^2={\rm inf} \frac{\int_{1}^\infty ds \left[-v^*v''+ \Lambda (1-\frac{1}{s^2})^2|v|^2\right]}{\int_1^\infty ds\left(\frac{s+1}{2s}\right)^4|v|^2}
\label{eq:sint}
\end{align}
Here we have made the substitutions $\rho\rightarrow r_0 s$, $(m_0r_0/4)^2\rightarrow\Lambda$.  Note that $s\geq 1$, and for the purposes of establishing a lower bound, we can replace the function $\left(\frac{s+1}{2s}\right)^4\rightarrow 1$ in the denominator of Eq.~(\ref{eq:sint}).

Now let us define a square well potential $U$ which is zero for all $s<s_0$ for some $s_0$, and $U=U_0=\Lambda(1-\frac{1}{s_0^2})^2$ for $s\geq s_0$. This potential is everywhere less than or equal to that of Eq.~(\ref{eq:sint}), so $r_0^2\omega^2$ is bounded by the ground state energy of the square well,
\begin{align}
r_0^2\omega^2\geq{\rm inf} \frac{\int_{1}^\infty ds \left[-v^*v''+U|v|^2\right]}{\int_1^\infty ds |v|^2}
\label{eq:lowerbound}
\end{align}

The ground state of the square well depends on the boundary condition at $s=1$. 
The quantum mechanical states of a relativistic scalar particle are obtained from classical solutions, so it is natural to impose boundary conditions that guarantee the conservation of the Noether charges associated with the symmetries of the background metric. Using the stress-energy tensor of a massless scalar field, we can see that this requirement implies that the following  fluxes vanish:
\begin{equation}
\label{energybc}
\mbox{$\oint$}_{\rho=r_0}\d x^{D-1}\sqrt{h}\,\partial_\ell\Phi\,\partial_\rho\Phi=0\;,
\end{equation}
where $h$ is the determinant of the metric induced on the sphere at $\rho=r_0$ and $\partial_\ell$ is either time derivative or differentiation with respect to the asimuthal angle to ensure the conservation of energy or angular momentum, respectively. It is sufficient to impose either the Dirichlet or the Neumann condition: $X=0$ or $\partial_\rho X=0$, where $\Phi(t,\cdot)=X(\cdot)e^{i\omega t}$. Therefore we accommodate general mixed boundary conditions by setting
\begin{align}
(\xi \partial_s v+ v)|_{s=1}=0
\end{align}
for arbitrary parameter $\xi$, 
which covers both the Dirichlet ($\xi=0$) and the Neumann ($\xi=-2$) condition. Since 
\begin{align}
\int_{s\geq1}\d x^D\,X^*(-\Delta X)&=\oint_{s=1}d x^{D-1}\sqrt{h_0}\,X^*\partial_rX+\int_{s\geq1}\d x^D\,\nabla X^*\cdot\nabla X\\
&= -\xi\oint_{s=1}d x^{D-1}\sqrt{h_0}\,|\partial_rX|^2+\int_{s\geq1}\d x^D\,\nabla X^*\cdot\nabla X,
\end{align}
where $\d x^{D-1}\sqrt{h_0}$ is the invariant measure induced by the $D$ dimensional Euclidean metric on the sphere at $s=\rho/r_0=1$. For the relevant values of $\xi$ ($\xi=0,-2$), the above expression is positive, and so is the potential $U$. Therefore the ground state energy of the square well is also positive, and we can seek the ground state inside the well in the form $e^{iks}+Be^{-iks}$, where $k^2\geq0$ (i.e.\ $k$ is real). Matching this function and its derivative with exponential decay $Ce^{-{\tilde k}s}$ ($\tilde k\equiv\sqrt{U_0-k^2}$) in the forbidden region, we obtain a constraint,
\begin{align}
\tan (k(s_0-1))=\frac{k \tilde k \xi-k}{k^2\xi+\tilde k}\;.
\end{align}
As usual, solving this constraint yields a discrete spectrum of energy levels $k^2$ for a given well size $s_0$.

Since we are interested in large $\Lambda$ scaling, we will take a convenient choice for $s_0$ and an ansatz for the behavior in this limit, then show that it yields solutions self-consistently.  Let us take $s_0=1+\beta/\Lambda^{1/4}$ for some $\beta$. This implies $U_0\sim4\beta^2\sqrt{\Lambda}$ for large $\Lambda$. We then try the scaling $k\rightarrow \alpha \Lambda^{1/4}$. For large $\Lambda$, the constraint simplifies and becomes $\Lambda$- and $\xi$-independent,
\begin{align}
\sqrt{-1+4\beta^2/\alpha^2}=\tan(\alpha\beta)\;.
\end{align}
Taking, for example, $\beta=1$, the solution for $\alpha$  is also $\calO(1)$. The ground state of the full potential therefore satisfies
\begin{align}
r_0^2\omega^2\geq \alpha^2\sqrt{\Lambda}\sim nT
\end{align}
for large $\Lambda$.

\bibliography{dipole_refs}{}
\bibliographystyle{utphys}

\end{document}